\documentclass{PoS}

\usepackage{graphicx}

\title{Interstellar and circumstellar fullerenes}

\ShortTitle{Interstellar and circumstellar fullerenes}

\author{\speaker{J. Bernard-Salas$^a$}, J. Cami$^{b,c}$, A.P. Jones$^d$, E. Peeters$^{b,c}$, E.R. Micelotta$^d$, M. Otsuka$^e$, G.C. Sloan$^f$, F. Kemper$^e$, M. Groenewegen$^g$\\
        $^a$Department of Physical Sciences, The Open University, Milton Keynes MK7 6AA, UK\\
        $^b$Department of Physics \& Astronomy, University of Western Ontario, Canada\\
	$^c$SETI Institute, Mountain View, CA 94043, USA\\
	$^d$Institut d'Astrophysique Spatiale, Orsay, France\\
	$^e$Academia Sinica, Institute for Astronomy \& Astrophysics, Taiwan\\
	$^f$Center for Radiophysics and Space Research, Cornell University, USA\\
	$^g$Royal Observatory of Belgium, Brussels, Belgium\\
        E-mail: \email{jeronimo.bernard-salas@open.ac.uk}}

\abstract{Fullerenes are a particularly stable class of carbon molecules in the shape of a hollow sphere or ellipsoid that might be formed in the outflows of carbon stars. Once injected into the interstellar medium (ISM), these stable species survive and are thus likely to be widespread in the Galaxy where they contribute to interstellar extinction, heating processes, and complex chemical reactions. In recent years, the fullerene species C$_{60}$ (and to a lesser extent C$_{70}$) have been detected in a wide variety of circumstellar and interstellar environments showing that when conditions are favourable, fullerenes are formed efficiently.
Fullerenes are the first and only large aromatics firmly identified in space. The detection of fullerenes is thus crucial to provide clues as to the key chemical pathways leading to the formation of large complex organic molecules in space, and offers a great diagnostic tool to describe the environment in which they reside. Since fullerenes share many physical properties with PAHs, understanding how fullerenes form, evolve and respond to their physical environment will yield important insights into one of the largest reservoirs of organic material in space. In spite of all these detections, many questions remain about precisely which members of the fullerene family are present in space, how they form and evolve, and what their excitation mechanism is. We present here an overview of what we know from astronomical observations of fullerenes in these different environments, and discuss current thinking about the excitation process. We highlight the various formation mechanisms that have been proposed, discuss the physical conditions conducive to the formation and/or detection of fullerenes in carbon stars, and their possible connection to PAHs, HACs and other dust features.
}

\FullConference{The Life Cycle of Dust in the Universe: Observations, Theory, and Laboratory Experiments - LCDU 2013,\\
		18-22 November 2013\\
		Taipei, Taiwan}

\begin{document}

\section{Introduction}

Fullerenes were discovered in
laboratory experiments aimed at understanding the formation of long
carbon chains in the circumstellar environment of carbon stars and
their survival in the interstellar medium (ISM) [1]. 
%
  Fullerenes can now be formed  efficiently
in laboratory experiments, converting a few percent of graphite into
C$_{60}$  [2, 3]. As soon as they were discovered, it was suggested that their extreme
stability, in particular against photodissocation, makes fullerenes
ideally suited to survive the harsh radiation field in the ISM [1]. There were many dedicated searches for fullerene signatures in interstellar, circumstellar and solar system environments. These searches did not, however, find conclusive evidence for the presence of these species (see [4] for a review).

Recently, using the {\em Spitzer Space Telescope}, Cami et al. [5] made the first conclusive detection of C$_{60}$ and C$_{70}$ in space, specifically, in the young planetary nebula Tc1. Following our discovery, C$_{60}$ has now been detected in many circumstellar and interstellar environments, including detections in many more evolved stars: planetary nebulae (PNe) in the Milky Way and the Magellanic Clouds [6, 7]; in
protoplanetary nebulae [8]; post-AGB stars [9]; in the surroundings of a few R CrB stars [10, 11] and in the
peculiar binary object XX Oph [12]. These observations
suggest that fullerenes are formed
by the complex and rich
circumstellar chemistry that occurs
in the short transition phase between 
AGB and PN phases. Moreover, the infrared C$_{60}$
bands have also been detected in the
interstellar medium. Following their
earlier suggestions [13], Sellgren et al. [14] confirmed
the presence of C$_{60}$ in the reflection nebulae NGC 7023 and NGC 2023, and showed that in NGC\,7023, the peak of fullerene emission comes from a different spatial location than that of the PAH bands. Rubin et al. [15] furthermore report the detection of C$_{60}$ in the Orion nebula, and Roberts et al. [16] found the C$_{60}$ bands in several young stellar objects and a Herbig Ae/Be star. These detections show that fullerenes can indeed survive the conditions in the ISM and become incorporated into the regions around young stars and, possibly, planetary systems.

Clearly, when conditions are favourable, fullerenes can form efficiently. The challenge resides now in understanding their excitation mechanisms and formation routes. Their excitation mechanisms are important to set their diagnostic value in the ISM. Their formation routes are crucial to understand the formation of complex organics in space.

\section{Excitation mechanisms}

Two different excitation mechanisms have been proposed for fullerene emission; thermal and fluorescent. Cami et al. [5] argued that the emission in Tc1 is consistent with thermal emission, whereas Sellgren et al. [14] interpreted the emission in NGC 7023 as infrared fluorescence resulting from stochastic heating. Although these excitation mechanisms result in different fullerene spectra, it has not been possible to conclusively favour one mechanism over the other by comparing the band intensities [17]. This is because of uncertainties in laboratory measurements of fullerene spectral properties and contamination by PAH features in many objects greatly complicate matters. In the infrared, C$_{60}$ has four active vibrational modes at 7.0, 8.5, 17.4 and 18.9\,$\mu$m. Observationally, most sources have a constant 18.9\,$\mu$m/17.4\,$\mu$m band ratio, which is difficult to reconcile with models of thermal emission, but it can be easily explained via fluorescence. On the other hand, fluorescence models predict a stronger band at 7\,$\mu$m than observed. 

The location and spatial variations of the fullerene emission can give important clues to solving this problem. In the case of thermal excitation, the distribution of fullerenes should closely follow the dust. The dust is predominately in thermal equilibrium with the central star's radiation field, and its temperature is determined by its location relative to the central star. If the fullerenes are emitting in thermal equilibrium, they should show similar variations with the distance to the star [17, 18]. In addition, if they are attached to dust grains as has been proposed, their temperature should be consistent with that expected for dust grains. Infrared fluorescence, on the other hand, should show a very weak dependence with distance to the central object. Using {\em Spitzer} observations Bernard-Salas et al. [17] show that the fullerene emission in Tc1 seems to arise 8000 AU from the central star (Figure 1, left panel).  High spatial resolution images from Gemini (Golriz et al., in prep.) have confirmed this distance, which is too far to explain the temperatures predicted by thermal emission ($\sim$300\,K in Tc1) and thus may favour fluorescence. Bern\'e et al. [19] have detected  the emission of ionised C$_{60}$ in NGC\,7023. They find that C$_{60}$$^+$ is emitting closer to NGC\,7023's excitation source (the binary system HD200775) than C$_{60}$, providing further support for fluorescence. It is however important to note that fluorescence models cannot fully reproduce the emission of the bands (overestimating the strength of the 7\,$\mu$m band), which may indicate that we are not seeing isolated fullerenes  but e.g. clusters. In addition to these mechanisms, excitation via collisions has not yet been explored in detail and should be further investigated. 

  \begin{figure}[]
  \begin{center}
   \includegraphics[width=6cm]{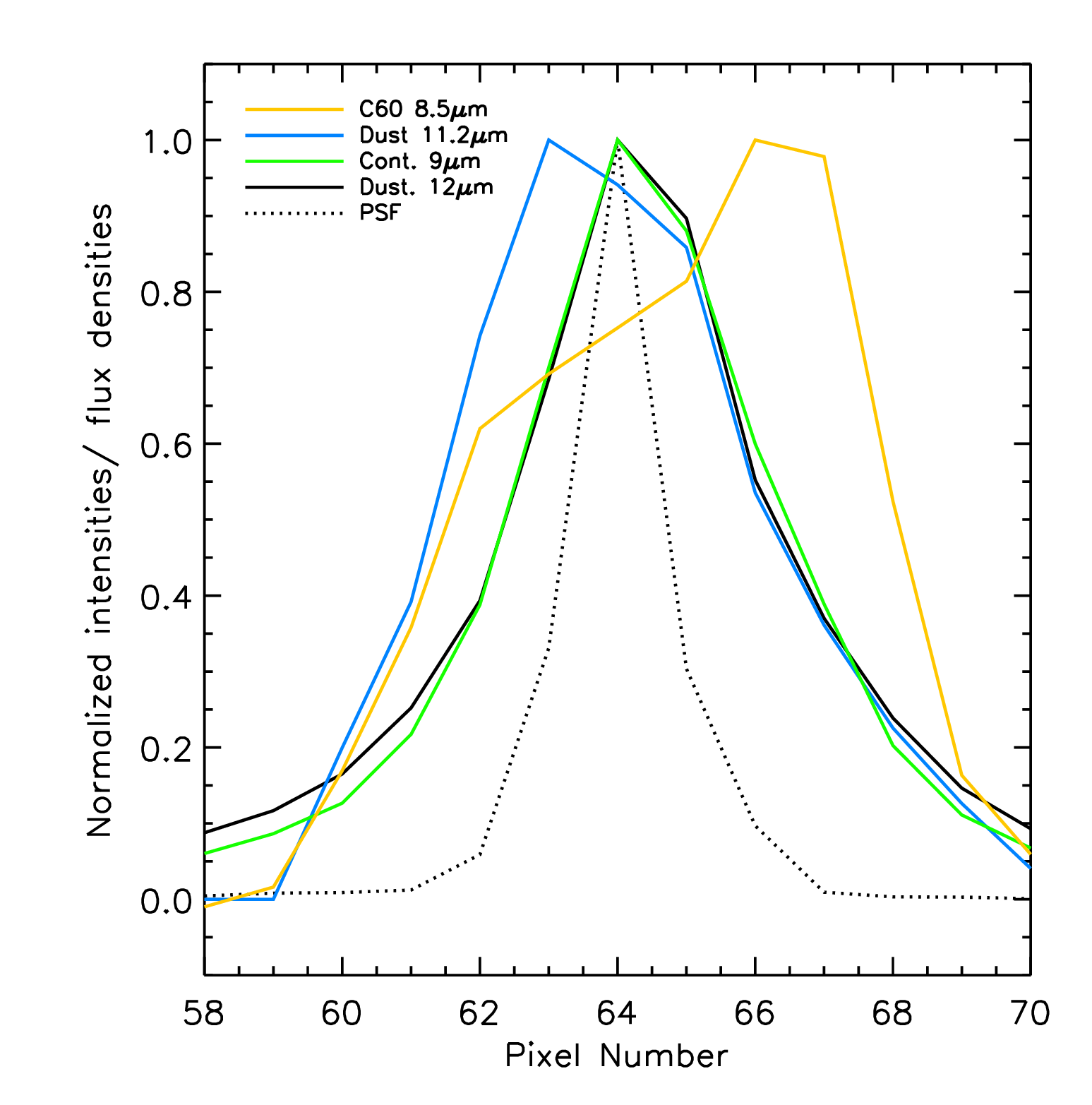}
   \includegraphics[width=8.2cm]{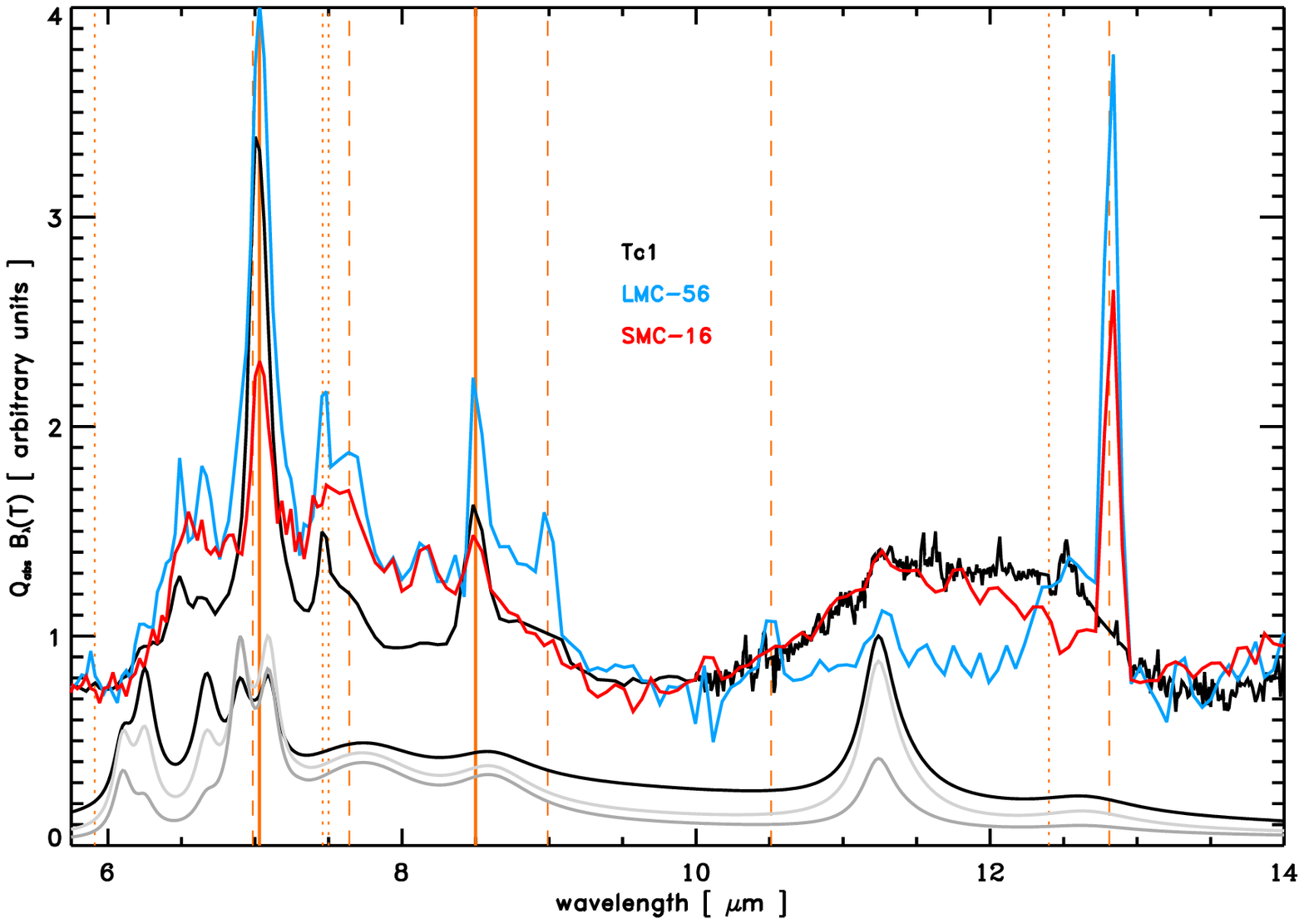}
 \end{center}
  \caption{{\em Left:}  Spatial distribution of different dust features in Tc1. The fullerene emission peaks far away from the central star and is not co-located with the PAH emission. {\em Right:} Comparison of the 6--9$\mu$m plateau of three PNe (in red, blue, and black) with a model spectra of HAC nanoparticles [17].}
\end{figure}

\section{Formation}

Over 60 molecular species and minerals, up to about 13 atoms in size, have been identified in the outflows of carbon-rich stars. A crucial step in the formation of the organic inventory needed for life is the formation of complex molecules.
With 60 and 70 carbon atoms, fullerenes
are now the largest molecules identified, strong evidence that large molecules can form in space. Their formation routes are thus important in understanding how large complex organic molecules can assemble in space. Several fullerene formation routes are possible in the laboratory or have been proposed, including: 1) low-temperature formation in the absence of hydrogen [1]; 2) high-temperature ($T>3000$ K)
formation in which case hydrogen may be present but PAHs are not formed [20]; 3) deformation of PAHs and graphene [19]; 4) photo-chemical processing of Hydrogenated Amorphous Carbon (HAC) [21]; and 5) formation via carbon vapour exposure in a process called closed network growth (CNG) [22, 33].

Micelotta et al. [24] have shown that  laboratory-based methods 1, 2, and 3 are not viable under astrophysical conditions. Methods 1 and 2 are based on coagulation and in space the density is too low for them to occur, while formation timescales from graphene are too long (method 3). Interestingly, sources with strong fullerene emission also exhibit a 6--9~$\mu$m plateau, probably related to their formation mechanism (Fig. 1). Bernard-Salas et al. [17] have shown that this plateau is consistent with the emission of HAC nanoparticles (Figure 1, right panel). This finding favours the formation of fullerenes via photo-chemical processing of HAC (method 4) but the mechanics of this evolutionary sequence are not yet fully understood. Based on this scenario, Micelotta et al. [24] propose a top-down mechanism of highly vibrationally-excited hydrocarbon cage-like nanoparticles that naturally results in their down-sizing to stable fullerene species. CNG is a formation mechanism based on laboratory experiments similar to the early experiments by Kroto et al. [1],  and show that fullerenes can grow upon direct exposure to carbon vapour. An interesting aspect of this mechanism is that it can trap other elements, resulting in metallofullerenes [23]. This breaks the symmetry of the molecule and activates silent modes which can produce emission in, for example, the 6--9~$\mu$m region. 

While we have not yet reached a consensus in the fullerene formation mechanism, evolution of HAC nanoparticles and CNG are potentially valid scenarios with interesting implications for the formation of complex organics. Any viable scenario, however, will need to be consistent with the fact that fullerenes are only seen in a small fraction of objects.

\section{Environment and carbon-rich evolution} 

The formation mechanism of fullerenes is key in understanding the evolution of carbon-rich dust. Being the first poly-aromatic species identified in circumstellar and interstellar environments, they provide a crucial anchor point to test models for dust evolution.   Otsuka et al. [25] and Sloan et al. [26] (see also proceedings for both authors [27, 28]) give  an overview of the physical properties in fullerene-rich sources and links between the different dust species in circumstellar environments.

The mid-infrared spectra of carbon-rich stars and PNe are very rich (e.g. [29]). Figure 2 (left) shows an example of Galactic fullerene containing planetary nebulae [25].  Sources with strong fullerene emission have little to no PAH emission, show strong plateau emission bands (e.g. 6--9 and 11--13\,$\mu$m), and always show the so-called 30\,$\mu$m feature. Sources with weaker fullerene bands present stronger PAHs, no plateaus, but still the 30$\mu$m feature. There is clearly an evolutionary or excitation link between the carriers of these features. The 30\,$\mu$m feature has been usually attributed to MgS [30]. However, Otsuka et al. [25] have shown that the 30\,$\mu$m correlates with the dust continuum, which they argue indicates a similar carrier of carbonaceous origin. It is also possible that the carriers differ, but both arise in similar conditions. Sloan et al. [26] have confirmed  that the broad 11\,$\mu$m feature seen in post-AGB and some PNe is SiC. 

\begin{figure}[]
  \begin{center}
   \includegraphics[width=8cm]{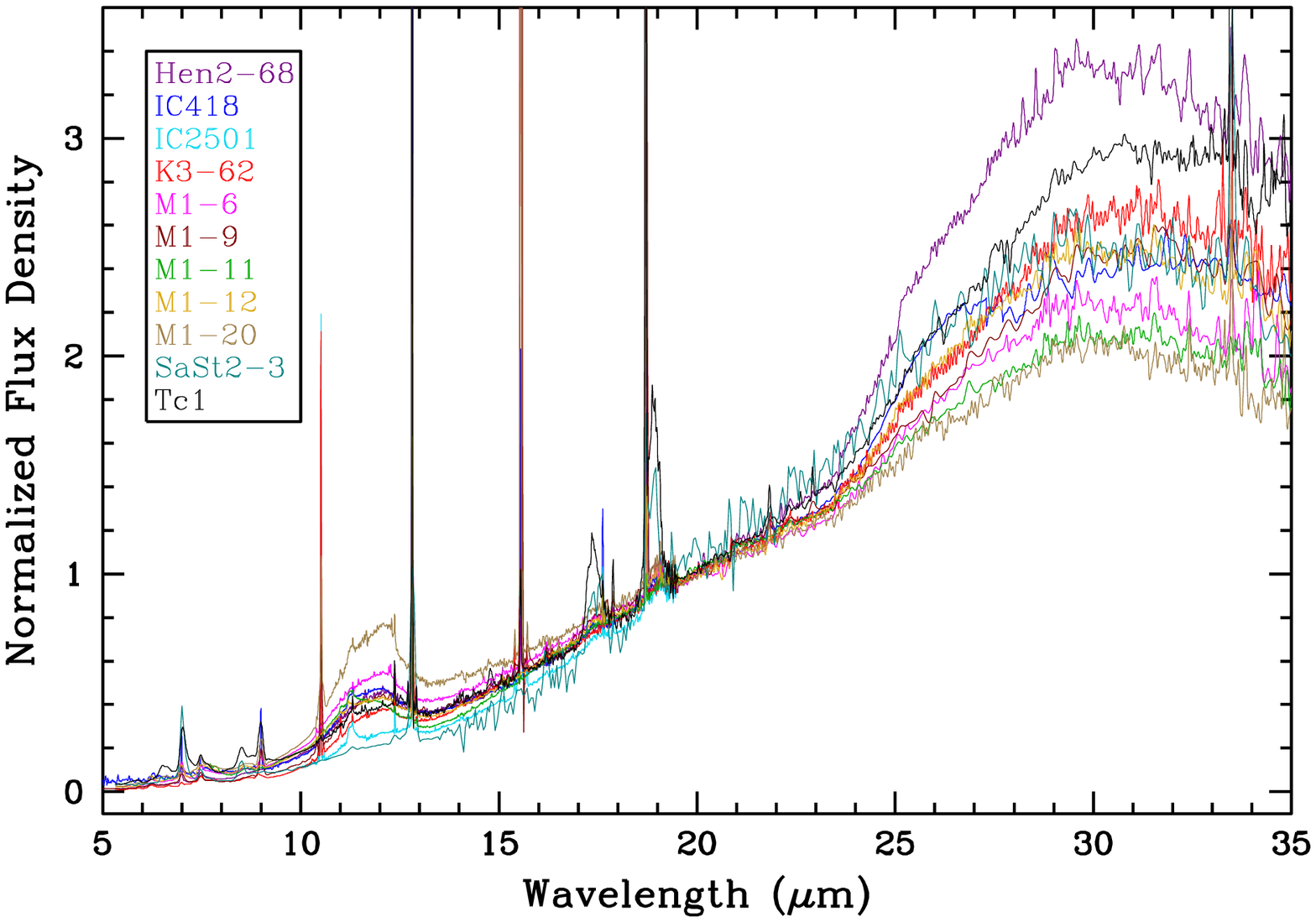}
   \includegraphics[width=6cm]{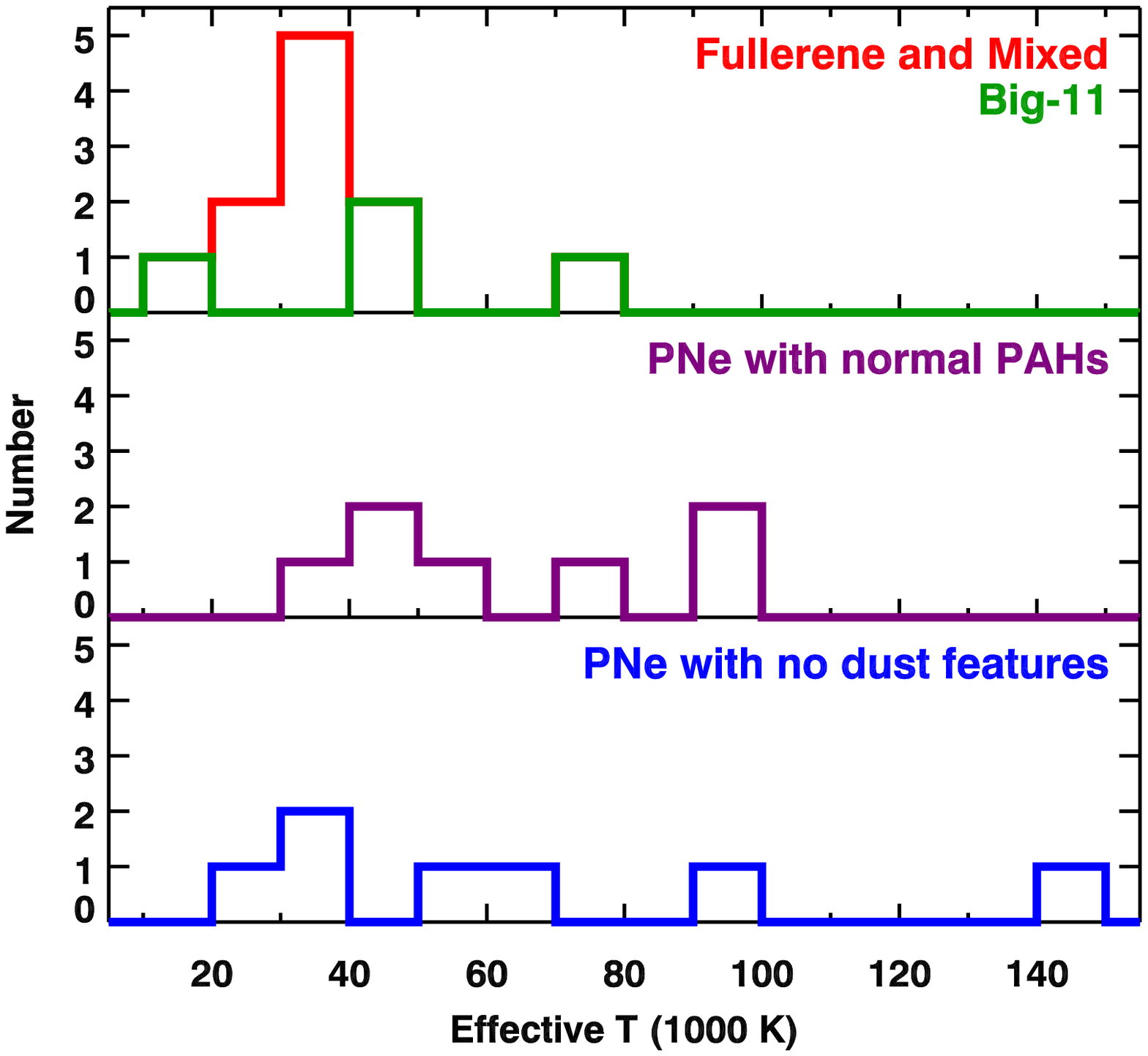}
 \end{center}
  \caption{{\em Left:} Continuum normalised spectra of several  PNe showing the diversity of features in the fullerene-rich sources(e.g. plateaus at 6--9\,$\mu$m and 10--13\,$\mu$m, and 30\,$\mu$m features) [25]. {\em Right:} Histogram showing the effective temperature of fullerene-containing carbon stars [26].}
\end{figure}

Finally, it is worth noticing that fullerenes are not common. They are detected in many diverse environments but they are only observed in 3\% of the Galactic carbon-rich PNe observed with {\em ISO} and/or {\em Spitzer}. Otsuka et al. [25] find that Galactic PNe with fullerene are located outside the solar circle, probably indicating a lower metallicity. Garcia-Hernandez et al. [31] and Sloan et al. [26] have shown that fullerenes do not require strong UV fields and are in fact seen in mild environments, with most circumstellar sources having effective temperatures of 30,000-40,000 K (see Figure 2, right panel). In addition, Sloan et al. [26] shows that objects with clear fullerene emission have optical bluer colours, probably indicating less circumstellar dust in the line of sight. Experiments by Dunk et al. [23] have shown that in the gas-phase  fullerene can react with PAHs,  providing insight into how fullerenes and metallo-fullerenes could enrich or be integrated into carbon grains, and thus linking molecular and solid state physics.

\section{Summary and concluding remarks}

Fullerenes have now been detected in many circumstellar and interstellar environments. With 60 and 70 carbon atoms, fullerenes are the largest molecules identified in space so far (and the first poly-aromatic species). This makes them crucial in our understanding the formation of complex organics that could later be incorporated into planetary systems, and provide key constraints into which anchor models for dust evolution.
The many detections in post-AGB stars, PNe and star forming environments indicate that fullerenes form in the last phase of evolution of carbon stars, and they can survive the conditions in the ISM. Their excitation mechanism is still being debated but current evidence seems to rule out thermal emission, and favours fluorescence. It is important to note that fluorescent models are not yet fully compatible with observations, indicating that maybe we are not seeing isolated free C$_{60}$ molecules. It is also possible that both emission mechanisms could be contributing, and which one dominates might depend on the physics of the source.
 Of the different formation mechanisms  that have been proposed, evolution of HACs nanoparticles and closed network growth could be the most viable scenarios, and have potentially interesting consequences for the  evolution of organic material. In the future it will be interesting to establish their role in extinction, their relevance connecting molecular to solid state physics, and exploring the chemistry occurring in their surfaces.  



\end{document}